\documentclass[10pt]{article}
\usepackage{graphicx}
\usepackage{amsmath}
\usepackage{amssymb}
\usepackage{caption2}
\begin{document}
\bibliographystyle{prsty}
\begin{center}
{\large {\bf \sc{ Strong decays of the charm mesons  $D_1^*(2680)$,  $D^*_3(2760)$,   $D_2^*(3000)$ }}} \\[2mm]
Zhi-Gang Wang  \footnote{E-mail:zgwang@aliyun.com. } \\
  Department of Physics, North China Electric Power University, Baoding 071003, P. R.
  China
\end{center}

\begin{abstract}
In this article, we assign   the higher charm mesons $D^*_1(2680)$, $D_3^*(2760)$ and $D_2^*(3000)$ to be the 2S $1^-$, 1D $3^-$ and 1F $2^+$ states, respectively, and study the two-body strong decays  to the ground state charm mesons and light pseudoscalar mesons with
the heavy meson  effective theory. We obtain the ratios among the strong decays, which can be confronted to the experimental
data in the future and shed light on the nature of those  higher charm mesons.
\end{abstract}

PACS numbers:  13.25.Ft; 14.40.Lb

{\bf{Key Words:}}  Charm mesons,  Strong decays
\section{Introduction}

Recently, the LHCb collaboration used the Dalitz plot analysis technique  to study the resonant substructures of $B^{-} \to D^{+} \pi^{-} \pi^{-}$ decays in a data sample corresponding to $3.0\, {\rm fb}^{-1}$ of $pp$ collision data recorded by the LHCb experiment during 2011 and 2012 \cite{LHCb1608}. A model-independent analysis of the angular moments indicated  the presence of resonances with spins 1, 2 and 3 at the $D^{+}\pi^{-}$ mass spectrum  \cite{LHCb1608}. The measured Breit-Wigner  masses and widths of those charm mesons are
\begin{flalign}
 & D_2^*(2460) : M = 2463.7 \pm 0.4 \pm 0.4 \pm 0.6 \mbox{ MeV}\, , \, \Gamma = 47.0\pm  0.8 \pm 0.9 \pm 0.3 \mbox{ MeV} \, , \nonumber\\
 & D_1^*(2680) : M = 2681.1 \pm 5.6 \pm 4.9 \pm 13.1 \mbox{ MeV}\, , \, \Gamma = 186.7 \pm 8.5 \pm 8.6 \pm 8.2 \mbox{ MeV} \, ,\nonumber \\
 & D^*_3(2760) : M = 2775.5 \pm 4.5 \pm 4.5 \pm 4.7 \mbox{ MeV} \, ,\, \Gamma = 95.3\pm  9.6 \pm 7.9 \pm 33.1 \mbox{ MeV} \, , \nonumber\\
 & D_2^*(3000) : M = 3214 \pm 29 \pm 33 \pm 36 \mbox{ MeV} \, ,\, \Gamma = 186 \pm 38 \pm 34 \pm 63 \mbox{ MeV} \, .
\end{flalign}
The $D_2^*(2460)$ is well established and the $J^P= 2^+$ assignment is strongly favored \cite{PDG}.
The mass and width of the $D_1^*(2680)$ state are close to those of the $D^*(2600)$  observed by the BaBar collaboration  \cite{Babar2010} and the $D_J^*(2650)$ observed by the  LHCb collaboration \cite{LHCb1307}.  The $D_1^*(2680)$, $D^*(2600)$ and  $D_J^*(2650)$ may be the same particle, and can assigned to be the 2S $1^-$ state \cite{Wang1009,Colangelo1207,D2600-2760,Wang1308,Godfrey1510}, see Table 1.

The mass and width of the $D^*_3(2760)^0$ state are close to those of the $D^*(2760)^0$ observed by the  BaBar collaboration  \cite{Babar2010} and the $D_J^*(2760)^0$ observed by the LHCb collaboration  \cite{LHCb1307}, and the  charged $D^*_3(2760)^+$ observed by the LHCb collaboration \cite{LHCb1505}. The $D^*_3(2760)^0$, $D^*(2760)^0$, $D_J^*(2760)^0$  may be the same particle, and can be assigned to be the 1D $3^-$ state \cite{Wang1009,Colangelo1207,D2600-2760,Wang1308,Godfrey1510,D2760}.
It is reasonable to assign  the $D^*_{3}(2760)$ to be the non-strange partner of the $D^*_{s3}(2860)$ according to the mass gap \cite{Wang1009,Colangelo1207,D2600-2760,Wang1308,Godfrey1510,D2760,Wang2860}, see Table 1.

 The $D^*_{sJ}(2860)$ meson was firstly observed  by  the BaBar  collaboration  in decays  to the final states  $D^0 K^+$ and $D^+K^0_S$ \cite{BaBar2006}, and
  confirmed by the BaBar  collaboration   in the decays to the final state $D^*K$  \cite{BaBar2009}.
 Later  the LHCb collaboration  observed a structure at $2.86\,\rm{GeV}$  in the $\overline{D}^0K^-$ mass spectrum in the Dalitz plot analysis of the decays $B_s^0\to \overline{D}^0K^-\pi^+$,  the structure contains both  the $D_{s1}^{*-}(2860)$ and the $D_{s3}^{*-}(2860)$ with $J^P=1^-$ and $3^-$, respectively \cite{LHCb7574,LHCb7712}. The Breit-Wigner masses and widths are
 \begin{flalign}
 & D_{s3}^*(2860) : M = 2860.5\pm 2.6 \pm 2.5\pm 6.0 \mbox{ MeV}\, , \, \Gamma = 53 \pm 7 \pm 4 \pm 6 \mbox{ MeV} \, , \nonumber\\
 & D_{s1}^*(2860) : M = 2859 \pm 12 \pm 6 \pm 23 \mbox{ MeV} \, ,\, \Gamma = 159 \pm 23\pm 27 \pm 72 \mbox{ MeV} \, .
\end{flalign}
 The energy gap  $ M_{D_{s3}^*(2860)}-M_{D_{3}^*(2760)}=85\,\rm{MeV}$,
 which is compatible with the $\overline{MS}$ mass $m_s(\mu=2\,\rm{GeV})=(95\pm5)\,\rm{MeV}$
 from the Particle Data Group \cite{PDG}. In the QCD sum rules for the $D_{s3}^*(2860)$, the optimal energy scale of the QCD spectral density is $\mu=2.1\,\rm{GeV}$ \cite{Wang1603}.   So it is reasonable to assign the $D^*_3(2760)$ to be the 1D $3^-$ state. In Ref.\cite{Godfrey1510},  Godfrey and  Moats also assign the $D_{s3}^*(2860)$ to be 1D $3^-$ state, although the value  from the relativized quark model or the Godfrey-Isgur model is $M_{1{\rm D},3^-}=2.917\,\rm{GeV}$.

The mass and width of the $D_2^*(3000)^0$  are not consistent with the    resonances  $D_J^*(3000)^0$ and $D_J^*(3000)^+$ observed previously by the LHCb collaboration \cite{LHCb1307},
\begin{flalign}
  & D_J^*(3000)^0 : M = 3008.1 \pm 4.0 \mbox{ MeV} \, ,\, \Gamma = 110.5 \pm11.5 \mbox{ MeV} \, , \nonumber\\
 & D_J^*(3000)^+ : M = 3008.1\, {\rm (fixed)}\, \mbox{ MeV} \, ,\, \Gamma = 110.5\, {\rm (fixed)}\,  \mbox{ MeV} \, .
\end{flalign}
The energy gap $M_{D_2^*(3000)^0}-M_{D_J^*(3000)^0}=206\,\rm{MeV}$, the $D_J^*(3000)^0$ and $D_2^*(3000)^0$ are different particles, see Table 1.
The strong decays
$D_J^*(3000)^0 \to D^{+}\pi^-$ and $D_J^*(3000)^+ \to D^{0}\pi^+$ were observed \cite{LHCb1307},
we can draw the  conclusion that the $D_J^*(3000)$ have the possible spin-parity $J^P=0^+,\,1^-,\,2^+,\,3^-,\,4^+,\cdots$. The  recent updated values of the masses of the 2P $0^+$ and $2^+$ states are $2.931\,\rm{GeV}$ and  $2.957\,\rm{GeV}$ respectively from the relativized quark model \cite{Godfrey1510}, we can tentatively assign the $D_J^*(3000)$ observed by the LHCb  collaboration to be the 2P $0^+$ or $2^+$ state, for detailed discussions about other possible assignments, one can consult Ref.\cite{Wang1308}.

If the  1D $2^-$ state $D^*_2$ and 1D $3^-$ state $D^*_3(2760)$ have approximately degenerate masses, then the energy gap
\begin{eqnarray}
M_{D_2^*(3000)}-M_{D^*_2}&=&438.5\,\rm{MeV}\, ,
\end{eqnarray}
the $D_2^*(3000)$ can be assigned to be the 2D  $2^-$ state. However, the decay $D_2^*(3000)\to D^+ \pi^-$ is  forbidden due to the conservation of parity.  According to the recent updated value of the mass of the 1F $2^+$ state $M=3.132\,\rm{GeV}$ from the relativized quark model \cite{Godfrey1510}, we can tentatively assign the $D_2^*(3000)$ to be the  1F $2^+$ state, the decays $D_2^*(3000) \to D^+\pi^-$ and $D^{*+}\pi^-$ can take place.

\begin{table}
\begin{center}
\begin{tabular}{|c|c|c|c|c|c|c|c|}\hline\hline
LHCb 2016 \cite{LHCb1608}   &BaBar 2010 \cite{Babar2010}   &LHCb 2013 \cite{LHCb1307}  &LHCb 2014 \cite{LHCb7574,LHCb7712} &$J^{P}$\\ \hline
$D_1^*(2680)^0/2681.1$      & $D^*(2600)^0/2608.7$         &$D_J^*(2650)^0/2649.2$     &                                   &$1^-$\\ \hline
$D_3^*(2760)^0/2775.5$      & $D^*(2760)^0/2763.3$         &$D_J^*(2760)^0/2760.1$     &                                   &$3^-$\\
                            &                              &                           &$D_{s3}^*(2860)/2860.5$            &$3^-$\\ \hline
$D_2^*(3000)^0/3214$        &                              &                           &                                   &$2^+$\\
                            &                              &$D_J^*(3000)^0/3008.1$     &                                   &$0^+,2^+$  \\ \hline
    \hline
\end{tabular}
\end{center}
\caption{ The  experimental values of the masses of the charm mesons, we present them in the form meson/mass, the unit of the mass is MeV. In the last column, we present the possible assignments of $J^P$.  }
\end{table}

In the article, we tentatively  assign   the higher charm mesons $D^*_1(2680)$, $D_3^*(2760)$ and $D_2^*(3000)$ to be the 2S $1^-$, 1D $3^-$ and 1F $2^+$ states, respectively, and study their two-body strong decays  with the heavy meson effective theory. Additional support can be obtained by the measuring the ratios among those strong decays. Charm meson spectroscopy  provides good opportunities to study QCD predictions based on the quark models. In the past years, there have been gained some new  experimental knowledge of the masses, widths and spins of the higher charm mesons and charm-strange mesons \cite{PDG}. The spectroscopic identification for the new higher states call for more experimental data and more theoretical works. In the present work, we will focus on the $D^*_1(2680)$, $D_3^*(2760)$ and $D_2^*(3000)$.

The article is arranged as follows:  we study the strong decays of the
$D_1^*(2680)$,  $D^*_3(2760)$,   $D_2^*(3000)$ with the heavy meson effective theory in Sect.2; in Sect.3, we present the
 numerical results and discussions; and Sect.4 is reserved for our
conclusions.

\section{ The strong  decays with the heavy meson effective theory }

 The  $c{\bar q}$ mesons  can be  sorted in doublets considering  the total
angular momentum of the light antiquark ${\vec s}_\ell$ in the heavy quark limit,
 where ${\vec s}_\ell= {\vec s}_{\bar q}+{\vec L} $,  the ${\vec
s}_{\bar q}$ and ${\vec L}$ are the light antiquark's  spin and orbital angular momentum, respectively \cite{RevWise}.
Now we write down  the spin-parity $J^P_{s_\ell}$ of the relevant doublets with $L=0,2,3$ explicitly,
\begin{eqnarray}
(D,D^*):         && (0^-,1^-)_{\frac{1}{2}} \, \, \, \, {\rm for} \,\,\,\, L=0 \, , \nonumber\\
(D^*_1,D_2):     && (1^-,2^-)_{\frac{3}{2}} \, \, \, \, {\rm for} \,\,\,\, L=2 \, , \nonumber\\
(D_2,D_3^{ *}):  && (2^-,3^-)_{\frac{5}{2}} \, \, \, \, {\rm for} \,\,\,\, L=2 \, , \nonumber\\
(D^*_2,D_3):     && (2^+,3^+)_{\frac{5}{2}} \, \, \, \, {\rm for} \,\,\,\, L=3 \, , \nonumber\\
(D_3,D_4^{ *}):  && (3^+,4^+)_{\frac{7}{2}} \, \, \, \, {\rm for} \,\,\,\, L=3 \, ,
\end{eqnarray}
 where the radial quantum numbers 1, 2, 3, $\cdots$ are not shown explicitly.
In the heavy meson effective theory,   the  spin doublets $(D,D^*)$,  $(D_2,D_3^{ *})$ and  $(D^*_2,D_3)$ can be
described by the  super-fields $H_a$,    $Y_a$ and $Z_a$,  respectively \cite{Falk1992},
\begin{eqnarray}
H_a & =& \frac{1+{\rlap{v}/}}{2}\left\{D_{a\mu}^*\gamma^\mu-D_a\gamma_5\right\} \, ,   \nonumber  \\
Y_a^{ \mu\nu} &=&\frac{1+{\rlap{v}/}}{2} \left\{D^{*\mu\nu\sigma}_{3a} \gamma_\sigma -D^{\alpha\beta}_{2a}\sqrt{5 \over 3} \gamma_5 \left[ g^\mu_\alpha g^\nu_\beta -{g^\nu_\beta\gamma_\alpha  (\gamma^\mu-v^\mu) \over 5} - {g^\mu_\alpha\gamma_\beta  (\gamma^\nu-v^\nu) \over 5}  \right]\right\}\, ,\nonumber\\
Z_a^{ \mu\nu} &=&\frac{1+{\rlap{v}/}}{2} \left\{D^{\mu\nu\sigma}_{3a}\gamma_5 \gamma_\sigma -D^{*\alpha\beta}_{2a}\sqrt{5 \over 3}  \left[ g^\mu_\alpha g^\nu_\beta -{g^\nu_\beta\gamma_\alpha  (\gamma^\mu+v^\mu) \over 5} - {g^\mu_\alpha\gamma_\beta  (\gamma^\nu+v^\nu) \over 5}  \right]\right\}\, ,
\end{eqnarray}
where the four vector $v_\mu$ satisfies $v^2=1$, the $a$ is the flavor index of the light antiquark, the  charm  meson fields  $D^{(*)}$ contain a factor $\sqrt{M_{D^{(*)}}}$ and
have dimension of mass $\frac{3}{2}$.

 The light pseudoscalar mesons are described by the fields
 $\displaystyle \xi=e^{i {\cal M} \over
f_\pi}$, where the matrix
\begin{equation}
{\cal M}= \left(\begin{array}{ccc}
\sqrt{\frac{1}{2}}\pi^0+\sqrt{\frac{1}{6}}\eta & \pi^+ & K^+\nonumber\\
\pi^- & -\sqrt{\frac{1}{2}}\pi^0+\sqrt{\frac{1}{6}}\eta & K^0\\
K^- & {\bar K}^0 &-\sqrt{\frac{2}{3}}\eta
\end{array}\right) \, ,
\end{equation}
and the decay constant $f_\pi=130\,\rm{MeV}$. In this article, we choose the definition $\langle 0|\bar{u}(0)\gamma_\alpha \gamma_5 d(0)|\pi(p)\rangle=if_\pi p_\alpha$. On the other hand, if we choose the definition $\langle 0|\bar{u}(0)\gamma_\alpha \gamma_5 d(0)|\pi(p)\rangle=i \sqrt{2}f_\pi p_\alpha$, then $f_\pi=92\,\rm{MeV}$.

We write down  the heavy meson chiral Lagrangians  ${\cal
L}_H$,  ${\cal L}_Y$ and ${\cal L}_Z$  describing
the strong decays to the  ground state charm mesons and light pseudoscalar mesons in the leading order approximation \cite{Wang1009,Wang1308,HL-1}:
\begin{eqnarray}
{\cal L}_H &=& \,  g_H {\rm Tr} \left\{{\bar H}_a H_b \gamma_\mu\gamma_5 {\cal A}_{ba}^\mu \right\} \, ,\nonumber \\
{\cal L}_{Y} &=&  {1 \over {\Lambda^2}}{\rm Tr}\left\{ {\bar H}_a Y^{\mu \nu}_b \left[k_1^Y \{{\cal D}_\mu, {\cal D}_\nu\} {\cal A}_\lambda + k_2^Y \left({\cal D}_\mu {\cal D}_\lambda { \cal A}_\nu + {\cal D}_\nu {\cal D}_\lambda { \cal A}_\mu \right)\right]_{ba}  \gamma^\lambda \gamma_5\right\} + h.c. \, , \nonumber\\
{\cal L}_{Z} &=&  {1 \over {\Lambda^2}}{\rm Tr}\left\{ {\bar H}_a Z^{\mu \nu}_b \left[k_1^Z \{ {\cal D}_\mu, {\cal D}_\nu\} {\cal A}_\lambda + k_2^Z \left({\cal D}_\mu {\cal D}_\lambda { \cal A}_\nu + {\cal D}_\nu {\cal D}_\lambda { \cal A}_\mu \right)\right]_{ba}  \gamma^\lambda \gamma_5\right\} + h.c. \, ,
\end{eqnarray}
where
\begin{eqnarray}
{\cal D}_{\mu}&=&\partial_\mu+{\cal V}_{\mu} \, , \nonumber \\
 {\cal V}_{\mu }&=&\frac{1}{2}\left(\xi^\dagger\partial_\mu \xi+\xi\partial_\mu \xi^\dagger\right)\, , \nonumber \\
 {\cal A}_{\mu }&=&\frac{1}{2}\left(\xi^\dagger\partial_\mu \xi-\xi\partial_\mu  \xi^\dagger\right)\,  , \nonumber\\
 \{ {\cal D}_\mu, {\cal D}_\nu \}&=&{\cal D}_\mu {\cal D}_\nu+{\cal D}_\nu {\cal D}_\mu  \, ,
\end{eqnarray}
the $g_H$,   $k^Y_1$, $k^Y_2$, $k^Z_1$ and $k^Z_2$ are  hadronic  coupling
constants, the $\Lambda$ is chiral symmetry-breaking energy scale and chosen as $ \Lambda = 1 \, \rm{GeV}$.

From the heavy meson chiral Lagrangians, we can obtain the partial decay  widths
$\Gamma$ for the two-body strong decays to the final states $D^*\mathcal{P}$ and $D\mathcal{P}$, where the $\mathcal{P}$ denotes the light pseudoscalar mesons \cite{Wang1308,WangEPJP},
\begin{eqnarray}
\Gamma\left[D_1^*(2680) \to D^*+{\mathcal{P}}\right] &=&C_{\mathcal{P}} \frac{g_H^2M_f p_f^3}{3\pi f_{\pi}^2M_i} \, , \\
\Gamma\left[D_1^*(2680) \to D+{\mathcal{P}}\right] &=&C_{\mathcal{P}} \frac{g_H^2M_f p_f^3}{6\pi f_{\pi}^2M_i} \, ,   \\
\Gamma\left[D_3^*(2760) \to D^*+{\mathcal{P}}\right] &=&C_{\mathcal{P}} \frac{16g_Y^2M_f p_f^7}{105\pi f_{\pi}^2\Lambda^4 M_i} \, , \\
\Gamma\left[D_3^*(2760) \to D+{\mathcal{P}}\right]   &=&C_{\mathcal{P}} \frac{4g_Y^2M_f p_f^7}{35\pi f_{\pi}^2\Lambda^4 M_i} \, , \\
\Gamma\left[D_2^*(3000) \to D^*+{\mathcal{P}}\right] &=&C_{\mathcal{P}} \frac{8g_Z^2M_f\left( p_f^2+m_\mathcal{P}^2\right) p_f^5}{75\pi f_{\pi}^2\Lambda^4 M_i } \, ,\\
\Gamma\left[D_2^*(3000) \to D+{\mathcal{P}}\right]   &=&C_{\mathcal{P}} \frac{4g_Z^2M_f \left(p_f^2+m_\mathcal{P}^2 \right) p_f^5}{25\pi f_{\pi}^2 \Lambda^4 M_i  } \, ,
\end{eqnarray}
where
\begin{eqnarray}
p_f&=&\frac{\sqrt{(M_i^2-(M_f+m_\mathcal{P})^2)(M_i^2-(M_f-m_\mathcal{P})^2)}}{2M_i}\, ,
\end{eqnarray}
  the $i$ and $f$ denote the initial and final state charm  mesons, respectively, $D^*=D^{*0},\,D^{*+},\,D_s^{*+}$, $D=D^0,\,D^+,\,D_s^+$, $g_Y=k^Y_1+k^Y_2$, $g_Z=k^Z_1+k^Z_2$.
 The coefficients $C_{\pi^\pm}=C_{K^\pm}=C_{K^0}=C_{\bar{K}^0}=1$, $C_{\pi^0}=\frac{1}{2}$ and $C_{\eta}=\frac{1}{6}$ or $\frac{2}{3}$. The values $C_{\eta}=\frac{1}{6}$ and $\frac{2}{3}$ correspond to the initial states $c\bar{u}$ (or $c\bar{d}$) and $c\bar{s}$, respectively.

\section{Numerical Results}
We take the masses of the light pseudoscalar mesons and the ground state charm mesons from the Particle Data Group,
$M_{\pi^+}=139.57\,\rm{MeV}$, $M_{\pi^0}=134.9766\,\rm{MeV}$,
$M_{K^+}=493.677\,\rm{MeV}$, $M_{\eta}=547.862\,\rm{MeV}$,
$M_{D^+}=1869.5\,\rm{MeV}$, $M_{D^0}=1864.84\,\rm{MeV}$,
$M_{D_s^+}=1969.0\,\rm{MeV}$, $M_{D^{*+}}=2010.27\,\rm{MeV}$,
$M_{D^{*0}}=2006.97\,\rm{MeV}$, $M_{D_s^{*+}}=2112.1\,\rm{MeV}$ \cite{PDG}.

Now we can obtain the partial decay widths from Eqs.(9-14), the numerical values are shown in Table 2, where we retain the hadronic coupling constants $g_H$, $g_Y$ and $g_Z$. The LHCb  collaboration measured the masses and widths of the
$D^*_1(2680)$, $D_3^*(2760)$, $D_2^*(3000)$, but did not measure the branching fractions of the  two-body strong decays $D^*_1(2680) \to D^+\pi^-$, $D_3^*(2760)\to D^+\pi^-$, $D_2^*(3000)\to D^+\pi^-$, we have no experimental data to fit the hadronic coupling constants $g_H$, $g_Y$ and $g_Z$. We can avoid the unknown   hadronic coupling constants $g_H$, $g_Y$ and $g_Z$ by studying the ratios $R_{D^*_1(2680)\to D^{(*)}\mathcal{P}}$, $R_{D^*_3(2760)\to D^{(*)}\mathcal{P}}$ and $R_{D_2^*(3000)\to D^{(*)}\mathcal{P}}$ among the strong decays of the  $D^*_1(2680)$, $D_3^*(2760)$ and $D_2^*(3000)$ mesons, respectively,
\begin{eqnarray}
R_{D^*_1(2680)\to D^{(*)}\mathcal{P}} &=&\frac{\Gamma\left[D^*_1(2680)\to D^{(*)}\mathcal{P}\right]}{\Gamma\left[D^*_1(2680)\to D^{*+}\pi^-\right]}\, , \nonumber\\
R_{D^*_3(2760)\to D^{(*)}\mathcal{P}} &=&\frac{\Gamma\left[D^*_3(2760)\to D^{(*)}\mathcal{P}\right]}{\Gamma\left[D^*_3(2760)\to D^{*+}\pi^-\right]}\, , \nonumber\\
R_{D_2^*(3000)\to D^{(*)}\mathcal{P}} &=&\frac{\Gamma\left[D_2^*(3000)\to D^{(*)}\mathcal{P}\right]}{\Gamma\left[D_2^*(3000)\to D^{*+}\pi^-\right]}\, .
\end{eqnarray}
In Table 3, we present the ratios $R_{D^*_1(2680)\to D^{(*)}\mathcal{P}}$, $R_{D^*_3(2760)\to D^{(*)}\mathcal{P}}$ and $R_{D_2^*(3000)\to D^{(*)}\mathcal{P}}$. By measuring those ratios, we can test the possible assignments and shed light on the nature of the  $D^*_1(2680)$, $D_3^*(2760)$, $D_2^*(3000)$ mesons.
The ratio between the kinematically allowed (or main) decays of the $D_2^*(2460)$ is 
 $\Gamma\left[D_2^*(2460)\to D^{+}\pi^-\right]/\Gamma\left[D_2^*(2460)\to D^{*+}\pi^-\right]=2.29$ from the heavy meson effective theory  in the leading order approximation \cite{Wang1308}, which differs from the ratios $R_{D^*_1(2680)\to D^{+}\pi^-}$, $R_{D^*_3(2760)\to D^{+}\pi^-}$ and $R_{D_2^*(3000)\to D^{+}\pi^-}$.
As far as  experimental  identifications of the charm mesons or charm-strange mesons are concerned, we can measure the mass spectra, angular momenta and parities of the $D^{* }\mathcal{P}$, $D\mathcal{P}$ systems to distinguish the $D_2^*(2460)$,  $D^*_1(2680)$, $D_3^*(2760)$, $D_2^*(3000)$, etc.

\begin{table}
\begin{center}
\begin{tabular}{|c|c|cc|cc| }\hline\hline
                 & $n\,L\,s_\ell\,J^P$         & Decay channels    & Widths  [GeV]      &Decay channels     & Widths [GeV]   \\ \hline

 $D^*_1(2680)$   & $2\,S\,\frac{1}{2}\,1^-$    & $D^{*+}\pi^-$     & $0.88938\,g_H^2$    & $D^{+}\pi^-$     & $0.68275\,g_H^2$   \\
                 &                             & $D^{*+}_sK^-$     & $0.07873\,g_H^2$    & $D_s^{+} K^-$    & $0.19979\,g_H^2$   \\
                 &                             & $D^{*0}\pi^0$     & $0.45189\,g_H^2$    & $D^{0}\pi^0$     & $0.34658\,g_H^2$   \\
                 &                             & $D^{*0}\eta$      & $0.03108\,g_H^2$    & $D^{0}\eta$      & $0.04806\,g_H^2$   \\  \hline

 $D_3^*(2760)$   & $1\,D\,\frac{5}{2}\,3^-$    & $D^{*+}\pi^-$     & $0.10016\,g_Y^2$    & $D^{+}\pi^-$     & $0.19128\,g_Y^2$        \\
                 &                             & $D_s^{*+}K^-$     & $0.00290\,g_Y^2$    & $D_s^{+}K^-$     & $0.02102\,g_Y^2$     \\
                 &                             & $D^{*0}\pi^0$     & $0.05174\,g_Y^2$    & $D^{0}\pi^0$     & $0.09884\,g_Y^2$      \\
                 &                             & $D^{*0}\eta$      & $0.00154\,g_Y^2$    & $D^{0}\eta$      & $0.00706\,g_Y^2$        \\ \hline

 $D_2^*(3000)$   & $1\,F\,\frac{5}{2}\,2^+$    & $D^{*+}\pi^-$     & $1.04657\,g_Z^2$    & $D^{+}\pi^-$     & $2.63140\,g_Z^2$        \\
                 &                             & $D_s^{*+}K^-$     & $0.42335\,g_Z^2$    & $D_s^{+}K^-$     & $1.30191\,g_Z^2$       \\
                 &                             & $D^{*0}\pi^0$     & $0.53129\,g_Z^2$    & $D^{0}\pi^0$     & $1.33823\,g_Z^2$      \\
                 &                             & $D^{*0}\eta$      & $0.10915\,g_Z^2$    & $D^{0}\eta$      & $0.30738\,g_Z^2$       \\ \hline \hline

\end{tabular}
\end{center}
\caption{ The strong decay widths of the three higher charm mesons with possible assignments.   }
\end{table}

\begin{table}
\begin{center}
\begin{tabular}{|c|c|c|c|c|c|c|c|c| }\hline\hline
                                        &$D^{*+}\pi^-$ &$D_s^{*+}K^-$ &$D^{*0}\pi^0$ &$D^{*0}\eta$ &$D^{+}\pi^-$ &$D_s^{+}K^-$ &$D^{0}\pi^0$ &$D^{0}\eta$ \\ \hline
$R_{D^*_1(2680)\to D^{(*)}\mathcal{P}}$ &1             &0.09          &0.51          &0.03         &0.77         &0.22         &0.39         &0.05  \\ \hline
$R_{D^*_3(2760)\to D^{(*)}\mathcal{P}}$ &1             &0.03          &0.52          &0.02         &1.91         &0.21         &0.99         &0.07  \\ \hline
$R_{D_2^*(3000)\to D^{(*)}\mathcal{P}}$ &1             &0.40          &0.51          &0.10         &2.51         &1.24         &1.28         &0.29 \\ \hline \hline
\end{tabular}
\end{center}
\caption{ The ratios $R_{D^*_1(2680)\to D^{(*)}\mathcal{P}}$, $R_{D^*_3(2760)\to D^{(*)}\mathcal{P}}$ and $R_{D_2^*(3000)\to D^{(*)}\mathcal{P}}$ among the strong decays of the three higher charm mesons.   }
\end{table}

\section{Conclusion}
In this article, we tentatively assign   the higher  charm mesons $D^*_1(2680)$, $D_3^*(2760)$ and $D_2^*(3000)$ to be the 2S $1^-$, 1D $3^-$ and 1F $2^+$ states, respectively, and resort to the heavy meson  effective Lagrangians in the leading order approximation to study their two-body strong decays
 to the ground state charm mesons and the light pseudoscalar mesons. We obtain the ratios $R_{D^*_1(2680)\to D^{(*)}\mathcal{P}}$, $R_{D^*_3(2760)\to D^{(*)}\mathcal{P}}$ and $R_{D_2^*(3000)\to D^{(*)}\mathcal{P}}$ among the strong decays of the  $D^*_1(2680)$, $D_3^*(2760)$, $D_2^*(3000)$ mesons, which can be confronted to the experimental
data in the future at the LHCb,  BESIII, KEK-B, and shed light on the nature of the  $D^*_1(2680)$, $D_3^*(2760)$, $D_2^*(3000)$ mesons.

\section*{Acknowledgement}
This  work is supported by National Natural Science Foundation,
Grant Number 11375063, and Natural Science Foundation of Hebei province, Grant Number A2014502017.

\end{document}